\documentclass[preprint,aps,showpacs]{revtex4}

\usepackage[dvips]{graphicx}

\usepackage{dcolumn}     % Align table columns on decimal point

\newcommand{\ftyf}{$^{45}/_{4}$}
\newcommand{\twnf}{$^{29}/_{4}$}
\newcommand{\twsf}{$^{27}/_{4}$}
\newcommand{\twff}{$^{25}/_{4}$}
\newcommand{\twtf}{$^{23}/_{4}$}
\newcommand{\twof}{$^{21}/_{4}$}
\newcommand{\nitf}{$^{19}/_{4}$}
\newcommand{\svtf}{$^{17}/_{4}$}
\newcommand{\fitf}{$^{15}/_{4}$}
\newcommand{\thtf}{$^{13}/_{4}$}
\newcommand{\elef}{$^{11}/_{4}$}
\newcommand{\ninf}{$^{9}/_{4}$}
\newcommand{\sevf}{$^{7}/_{4}$}
\newcommand{\fivf}{$^{5}/_{4}$}
\newcommand{\thrf}{$^{3}/_{4}$}
\newcommand{\onef}{$^{1}/_{4}$}
\newcommand{\twsh}{$^{27}/_{2}$}
\newcommand{\twth}{$^{23}/_{2}$}
\newcommand{\twoh}{$^{21}/_{2}$}
\newcommand{\svth}{$^{17}/_{2}$}
\newcommand{\fith}{$^{15}/_{2}$}
\newcommand{\thth}{$^{13}/_{2}$}
\newcommand{\eleh}{$^{11}/_{2}$}
\newcommand{\ninh}{$^{9}/_{2}$}
\newcommand{\sevh}{$^{7}/_{2}$}
\newcommand{\fivh}{$^{5}/_{2}$}
\newcommand{\thrh}{$^{3}/_{2}$}
\newcommand{\oneh}{$^{1}/_{2}$}

\begin{document}
 
{

\title{Pair counting, pion-exchange forces, and the structure of light nuclei}

\author{R. B. Wiringa\cite{rbw}}

\affiliation{Physics Division, Argonne National Laboratory, 
             Argonne, Illinois 60439}
\date{\today}
 
\begin{abstract}
A simple but useful guide for understanding the structure of light nuclei
is presented.  It is based on counting the number of interacting pairs in
different spin-isospin ($S,T$) states for a given spatial symmetry, and 
estimating the overall binding according to the sum of 
${\bf\sigma}_{i}\cdot{\bf\sigma}_{j} {\bf\tau}_{i}\cdot{\bf\tau}_{j}$
expectation values, as suggested by one-pion-exchange.  Applied to s- and 
p-shell nuclei, this simple picture accounts for the relative stability of
nuclei as $A$ increases and as $T$ changes across isobars, the saturation
of nuclear binding in the p-shell, and the tendency to form $d$, $t$, or
$\alpha$ subclusters there.  With allowance for pairwise tensor and
spin-orbit forces, which are also generated or boosted by pion-exchange,
the model explains why mixing of different spatial symmetries in ground
states increases as $T$ increases across isobars, and why for states of
the same spatial symmetry, the ones with greater $S$ are lower in the 
spectrum.  The ordering of some sd-shell intruder levels can also be 
understood.  The success of this simple model supports the idea that 
one-pion-exchange is the dominant force controlling the structure of light
nuclei.
\end{abstract}
 

\maketitle

}

\section {Introduction}

The last decade has seen significant progress both in the characterization
of realistic two- and three-nucleon interactions, and in the ability to
make accurate many-body calculations with these models.  Nucleon-nucleon
potentials such as Argonne $v_{18}$ \cite{WSS95}, CD-Bonn \cite{MSS96,MB01},
and the Nijmegen models \cite{SKTS94} reproduce $N\!N$ scattering data
extremely well; when combined with three-nucleon forces such as the
Illinois or Tucson-Melbourne potentials \cite{PPWC01,CH01} and accurate 
many-body techniques, nuclear binding energies up to $A=12$ can be reproduced.
Seven different many-body methods are in superb agreement for the
binding energy of $^4$He with a realistic $N\!N$ force \cite{KNG01}, while
Green's function Monte Carlo (GFMC) \cite{PW01,PVW02,PWC04,P05},
no-core shell model (NCSM) \cite{NVB00,CNOV02,NO03}, and coupled-cluster
methods (CCM) \cite{KDHPP04,WDGHKPP05} are making very successful 
{\it ab initio} calculations for p-shell nuclei.  This progress
allows us to study the interplay between nuclear forces and nuclear 
structure in an unprecedented manner.

In a recent letter \cite{WP02} we constructed a series of increasingly 
realistic force models and used GFMC calculations to evaluate the
consequences for nuclear structure.  This study showed that a simple
central potential, with the canonical intermediate-range attraction and
short-range repulsion indicated by $S$-wave $N\!N$ phase shifts, could
approximately reproduce the triton and alpha binding energies, but failed 
to saturate in the p-shell, producing stable $^5$He and greatly overbinding
the $A$=6,7,8 nuclei.  To obtain unstable $^5$He and the general saturation
of nuclear forces that is evident in the p-shell, it is necessary to have
a state-dependent force, i.e., one that is attractive in $L$=even partial 
waves and repulsive in $L$=odd partial waves.  Indeed, state-dependence
appears to be more important for nuclear saturation than either the repulsive
core or the finite range of nuclear forces.  To obtain the further 
refinement that $^8$Be is unstable against breakup into two alphas requires
the addition of a tensor force, while the stability of $^{6,7}$Li suggests
spin-orbit terms are also needed.  Together, these are the major operator
components required in a realistic interaction to fit $S$- and $P$-wave 
$N\!N$ data.

Pion-exchange forces play a very significant role here.  The spin-isospin 
dependence of one-pion exchange (OPE) makes it attractive in $L$=even partial
waves and repulsive in $L$=odd partial waves, just as required to bind the 
s-shell nuclei but saturate quickly in the p-shell.  OPE also is the major
source of the tensor force, and iterated tensor interactions between
three or more nucleons provide a large enhancement to spin-orbit splitting in
nuclei \cite{PP93}.  In GFMC calculations of $A \leq 12$ nuclei with 
realistic interactions, the expectation value of the OPE potential is 
typically 70--75\% of the total potential energy \cite{PW01}.
The importance of pion-exchange forces is even greater when one considers
that much of the intermediate-range attraction in the $N\!N$ interaction
can be attributed to uncorrelated two-pion exchange with the excitation of 
intermediate $\Delta(1232)$ resonances \cite{WSA84}.  In addition,
two-pion exchange between three nucleons is the leading term in $3N$
interactions, which are required to get the empirical binding in
light nuclei \cite{PPWC01}.  In particular, the $3N$ forces provide the extra
binding required to stabilize the Borromean nuclei $^{6,8}$He and $^9$Be.

The thesis of this paper is that by counting the number of different 
spin-isospin ($S,T$) pairs that occur in a given nuclear state of specific 
spatial symmetry, and multiplying by a numeric strength taken from the OPE 
operator ${\bf\sigma}_{i}\cdot{\bf\sigma}_{j} {\bf\tau}_{i}\cdot{\bf\tau}_{j}$,
a very good measure of the binding energy is obtained.  This works both for 
the relative energy between different states in the same nucleus, and 
between different nuclei.
The idea is akin to the supermultiplet theory of Wigner \cite{EW58} which 
focused on the symmetry aspects of light nuclei, but assumed forces that 
were primarily central and space-exchange in character.  The present study
benefits from the extensive recent progress in fully realistic calculations 
mentioned above.
There is also common ground with the recent work by Otsuka and collaborators
\cite{OFUBHM01,OSFGA05} within the framework of traditional shell model
that emphasizes the importance of OPE spin-isospin and tensor interactions
in determining how single-particle energy levels shift as shells are filled.

This simple guide, supplemented by our knowledge of tensor, spin-orbit, 
and Coulomb forces, describes the general structure of light nuclei in 
considerable detail.  The model explains the growing binding as $A$ increases,
the saturation of binding going from the s-shell to the p-shell, the relative 
stability as $T$ varies across isobars, and the tendency to form $d$, $t$, 
and $\alpha$ subclusters in the light nuclei.  It explains why
mixing of different spatial symmetries in ground states increases as $T$
increases and why for states of the same spatial symmetry, the ones of 
higher $S$ are lower in the spectrum.  The same logic can also be used to 
understand the ordering of some sd-shell intruder levels in some of these 
nuclei.  

\section {Pair counting}

The total number of pairs in a nucleus, $P_A=A(A-1)/2$, can be subdivided 
into pairs of specific spin and isospin $P_A(ST)$ where $S=0$ or 1 and $T=0$ 
or 1.  Starting with the square of the expression 
$\sum_i{{\bf\tau}_{i}/2} = T_A$, where $T_A$ is the total isospin of the 
nucleus, and using the projection operators 
$(1-{\bf\tau}_{i}\cdot{\bf\tau}_{j})/4$ and
$(3+{\bf\tau}_{i}\cdot{\bf\tau}_{j})/4$ for $T=0$ and 1 pairs, respectively,
the total number of such pairs of given isospin in a nucleus can be shown 
to depend only on $A$ and $T_A$~\cite{FPPWSA96}:
\begin{eqnarray}
P_A(10) + P_A(00) = \frac{1}{8} [~~A^2 + 2A - 4T_A(T_A+1)~] \ , \\
P_A(11) + P_A(01) = \frac{1}{8} [~3A^2 - 6A + 4T_A(T_A+1)~] \ .
\end{eqnarray}
A similar pair of equations can be obtained for the total number
of $S=0$ or 1 pairs in terms of the total nuclear spin $S_A$, 
assuming spin is conserved, i.e., before configuration mixing by tensor 
forces and correlations.  An additional expression can be obtained for the
difference, $p_{[n]}$, between the number of symmetric (even) and 
antisymmetric (odd) pairs for a given spatial symmetry state specified 
by the Young diagram $[n]$:
\begin{equation}
P_A(10) + P_A(01) - P_A(11) - P_A(00) = p_{[n]} \ .
\end{equation}
For example a [3] symmetry state has three symmetric pairs, $p_{[3]}=3$; 
a [111] state has three antisymmetric pairs, $p_{[111]}=-3$; and a [21] state 
has one symmetric, one antisymmetric, and one mixed symmetry pair (which
does not contribute here), giving $p_{[21]}=0$.  Together we have four 
independent relations for four unknowns, which can be rearranged to give:
\begin{eqnarray}
\label{eq:p11}
P_A(11) &=& \frac{1}{4}[ ~2P_A -p_{[n]}-\frac{3}{2}A+S_A(S_A+1)+T_A(T_A+1)~] \ , \\
P_A(10) &=& \frac{1}{4}[ ~~P_A +p_{[n]}     ~~~~~~~~+S_A(S_A+1)-T_A(T_A+1)~] \ , \\
P_A(01) &=& \frac{1}{4}[ ~~P_A +p_{[n]}     ~~~~~~~~-S_A(S_A+1)+T_A(T_A+1)~] \ , \\
P_A(00) &=& \frac{1}{4}[~~~~~~ -p_{[n]}+\frac{3}{2}A-S_A(S_A+1)-T_A(T_A+1)~] \ .
\label{eq:p00}
\end{eqnarray}

The simple energy measure we propose using is obtained by multiplying the
number of pairs of each type with the expectation value of the spin-isospin
operator ${\bf\sigma}_{i}\cdot{\bf\sigma}_{j} {\bf\tau}_{i}\cdot{\bf\tau}_{j}$ 
coming from one pion-exchange:
\begin{equation}
E_{OPE} = C~[~P_A(11) - 3 P_A(10) -3 P_A(01) + 9 P_A(00)~] \ ,
\label{eq:ope}
\end{equation}
where $C$ is a constant in MeV.  
A value $C \sim 1.5$ MeV gives a reasonably good average scale factor.
This expression reflects the fact that $S$-wave $N\!N$ interactions are 
attractive while $P$-wave interactions are repulsive.  It does not attempt 
to differentiate between $^1S_0$ and $^3S_1$-$^3D_1$ channels, when in 
reality the former is just unbound and the latter produces a bound deuteron, 
thanks largely to the OPE tensor force.  However it does reflect the large 
difference between the weakly repulsive $^3P_J$ channels and the strongly 
repulsive $^1P_1$ interaction.  We will find that this simple expression 
does a remarkably good job of predicting overall trends in
binding and relative stability for s- and p-shell nuclei, as well as 
explaining a variety of observed features in the excitation spectra.

\section {Energy Spectra}

The $P_A(ST)$ and $E_{OPE}$ for $A$=2--5 nuclei are shown in 
Table~\ref{tab:a25}.  (In the following tables, we will only show the most
neutron-rich member of any isobaric multiplet, e.g. $^3$H but not $^3$He,
and $^6$He but not $^6$Be or the isobaric analog states in $^6$Li; they
should be understood to be essentially the same for nuclear forces and
differ primarily by the Coulomb energy.)  The deuteron, $^2$H, has
only one $ST$=10 pair, which we assign the strength $-3C$.
In our simple model, the $ST$=01 dineutron would also be bound, whereas in
reality it is just unbound, and is not shown.  The triton, $^3$H, has 
three pairs, equally divided between $ST$=10 and 01 according to 
Eqs.(\ref{eq:p11}-\ref{eq:p00}), and thus gets a strength of $-9C$, while 
the alpha, $^4$He, has six such pairs with total strength of $-18C$.  

If we use $E_{OPE}$ to judge the relative binding of these nuclei, then the
$d$~:~$t$~:~$\alpha$ energies should be in the ratio 1~:~3~:~6, whereas in 
reality they are more like 1~:~4~:~13.  Of course the binding is the result 
of a cancellation between kinetic and potential energies, and $E_{OPE}$ is
essentially a potential measure.  In fact, GFMC calculations for the
AV18/IL2 Hamiltonian give the expectation values for the two-body potential 
to be in the ratio 1~:~2.7~:~6.3 for these nuclei \cite{PPWC01}, reasonably 
close to $E_{OPE}$.  The $E_{OPE}$ will be a useful gauge for binding energies
only if there is something like a virial theorem for nuclei that says the
kinetic and potential energies are proportional to each other.  Fortunately
there does seem to be such a relation, at least in the light p-shell nuclei, as
shown by the results of GFMC calculations displayed in Table~\ref{tab:virial}.
For $6 \leq A \leq 12$ nuclei, the ratio, $R_{KV}$, of kinetic to potential 
energy expectation values varies only from 0.76 to 0.81, while there is a much
greater range of 0.75 to 0.90 for the s-shell nuclei.  We note that the 
lowest ratios occur for the most spatially-symmetric nuclei $^4$He, $^8$Be 
and $^{12}$C as one might expect.

A further complication is that when realistic tensor forces are included, 
some fraction of the $ST$=01 pairs will be converted to 11 pairs, and a 
small fraction of $ST$=10 pairs to 00 pairs, due to multi-body correlations.
Variational Monte Carlo (VMC) calculations for $^4$He found the actual 
distribution of 11~:~10~:~01~:~00 pairs to be 0.47~:~2.53~:~2.99~:~0.01 
\cite{FPPWSA96}.  For our simple model we will focus on the distribution 
of pairs before such mixing takes place.

While this very qualitative model is not particularly useful for the s-shell, 
it starts to have some utility in the p-shell.  The ten pairs of nucleons
in $^5$He are divided into six pairs within the s-shell, designated ``ss''
in Table~\ref{tab:a25}, and four pairs with one nucleon in the s-shell and
one in the p-shell, designated ``sp''.  The ss pairs are distributed exactly
as in $^4$He, while the sp pairs come in just the right combination to give
no additional contribution to $E_{OPE}$.  Thus the prediction is that $^5$He
should have the same binding as $^4$He, when in fact it is unstable against
breakup by $\sim$1 MeV.  For cases like this, our simple measure is not
sufficient to determine stability, but only to indicate a situation that 
could go either way, depending on, for example, how much the virial ratio, 
$R_{KV}$, of Table~\ref{tab:virial} varies.  In the case of $^5$He,
the residual attraction from shorter-range $N\!N$ and $3N$ forces is not
enough to overcome the additional kinetic energy that is generated by the
requirement of putting the fifth nucleon in a p-shell orbital.

For $A$=6 nuclei, shown in Table~\ref{tab:a6}, the s-shell core remains
the same, while the number of sp pairs doubles, but still with no net
contribution to $E_{OPE}$.  Effectively, the s- and p-shells decouple from
each other at the OPE level, which may help us understand why the use of
an inert core in the traditional shell model approach is valid.  The
final energy of $^6$Li and $^6$He then depends on the last pair of nucleons
which is wholly within the p-shell, designated ``pp'' in Table~\ref{tab:a6}.
Starting with $A$=6, there are multiple ways of adding up orbital and spin 
angular momenta to get the total $J^\pi;T$ of a given nuclear state 
\cite{BM69}; we label them by their $LS$ coupling and spatial symmetry, 
$^{2S+1}L[n]$, and list all allowed $L$ values.  For $^6$Li this last 
pair can be part of a $^3S$[42] or $^3D$[42] state (essentially a deuteron 
with orbital momentum of 0 or 2 around an alpha core) with an associated 
strength of $-3C$, or the last pair can be part of a $^1P$[411] state, which 
contributes $+9C$ to our binding measure.  The total $E_{OPE}$ for $^6$Li 
is the sum of the ss and pp pairs, or $-21C$ for the [42] states and $-9C$ 
for the [411] state.  For $^6$He the last pair can be part of either a
$^1S$[42] or $^1D$[42] state (essentially a spin zero dineutron with 
$L=0$ or 2 around an alpha core) with strength $-3C$, or part of a
$^3P$[411] state with strength $+C$; the corresponding total $E_{OPE}$ 
is $-21C$ or $-17C$, respectively.  

\begin{figure}[ht!]
\centering
\includegraphics[width=4.8in]{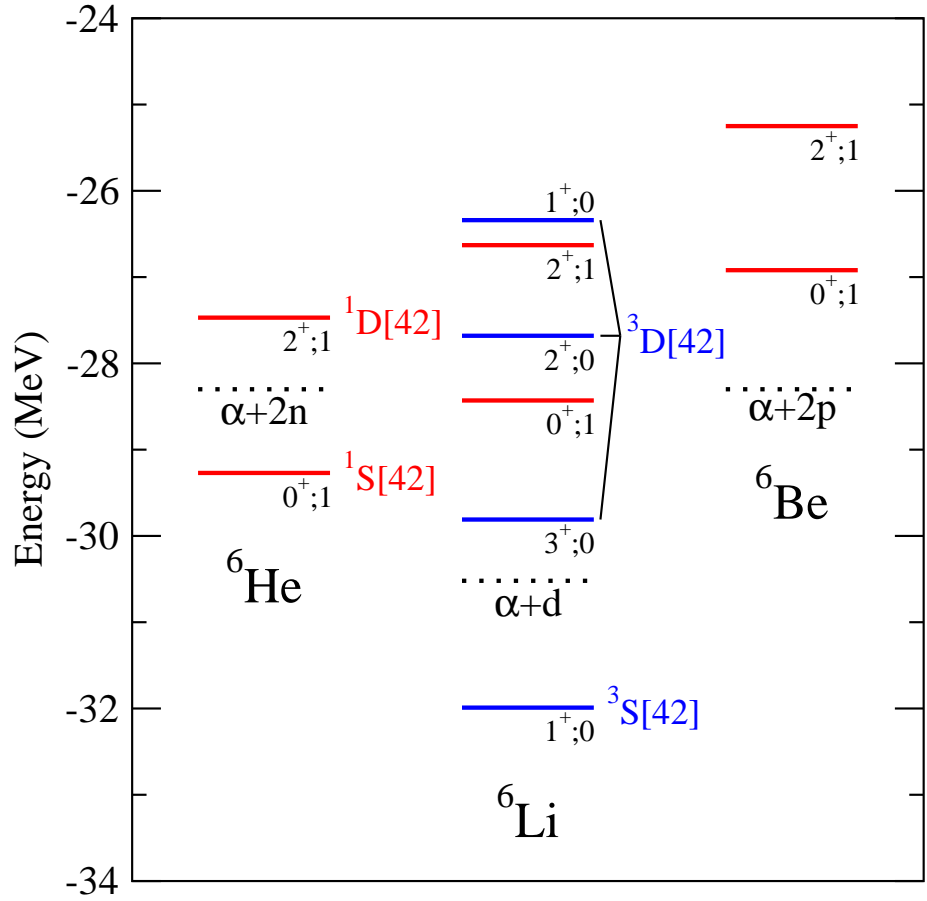}
\caption{(Color online) Experimental spectrum for $A=6$ nuclei; $T=0$ ($T=1$)
states are shown by blue (red) solid lines and breakup thresholds by black
dotted lines.}
\label{fig:a6}
\end{figure}

Thus the prediction of our simple model is that $^6$Li and $^6$He ground
states should have [42] symmetry, with about the same energy, and be weakly 
bound compared to $^4$He, which is pretty much correct, given the above 
caveat about unbound dineutron and bound deuteron.  The experimental spectrum 
\cite{exp567} is shown in Fig.~\ref{fig:a6}, where the levels are labeled 
by their dominant symmetry.  The $^6$Li ground state is 1.47 MeV below the 
alpha-deuteron threshold, while $^6$He is 0.97 MeV below the alpha-dineutron 
threshold.  GFMC calculations indicate that much of the binding between 
clusters is provided by the $3N$ force; if only the AV18 $N\!N$ force is used,
$^6$Li is stable by 0.6 MeV and $^6$He is unstable by 0.3 MeV \cite{PWC04}.

Not surprisingly, in the excitation spectrum the $D$ states are higher than 
the $S$ states, because the angular momentum barrier screens some of the
overall potential attraction.  In shell model studies this feature is
taken into account by including an $L^2$ term in the interaction \cite{M01}, 
but the spread between different $L$-states of the same spatial symmetry
is generally smaller than the spacing between different spatial symmetry
groups, and here we are after only the most general nuclear structure aspects.
Further, the $^3D$ combination in $^6$Li is split into $J$=1,2,3 states 
ordered with maximum $J$ lowest, as dictated by the spin-orbit force.
The antisymmetric [411] states are several MeV higher in the spectrum 
and no corresponding experimental states have been identified.  
However our simple model predicts that the antisymmetric $^1P$[411] 
state in $^6$Li is much higher than the $^3P$[411] state in $^6$He,
which suggests that when configuration mixing with tensor forces
is done, the admixture of these components in the respective ground 
states will be less for $^6$Li than for $^6$He.  This is borne out in
VMC diagonalizations with realistic forces where the amplitudes of the
different components in the ground state are 0.98~:~0.14~:~0.10 for
the $^3S$[42]~:~$^3D$[42]~:~$^1P$[411] pieces in $^6$Li and 0.97~:~0.23 for
the $^1S$[42]~:~$^3P$[411] pieces in $^6$He \cite{PWC04}.

\begin{figure}[ht!]
\centering
\includegraphics[width=6.0in]{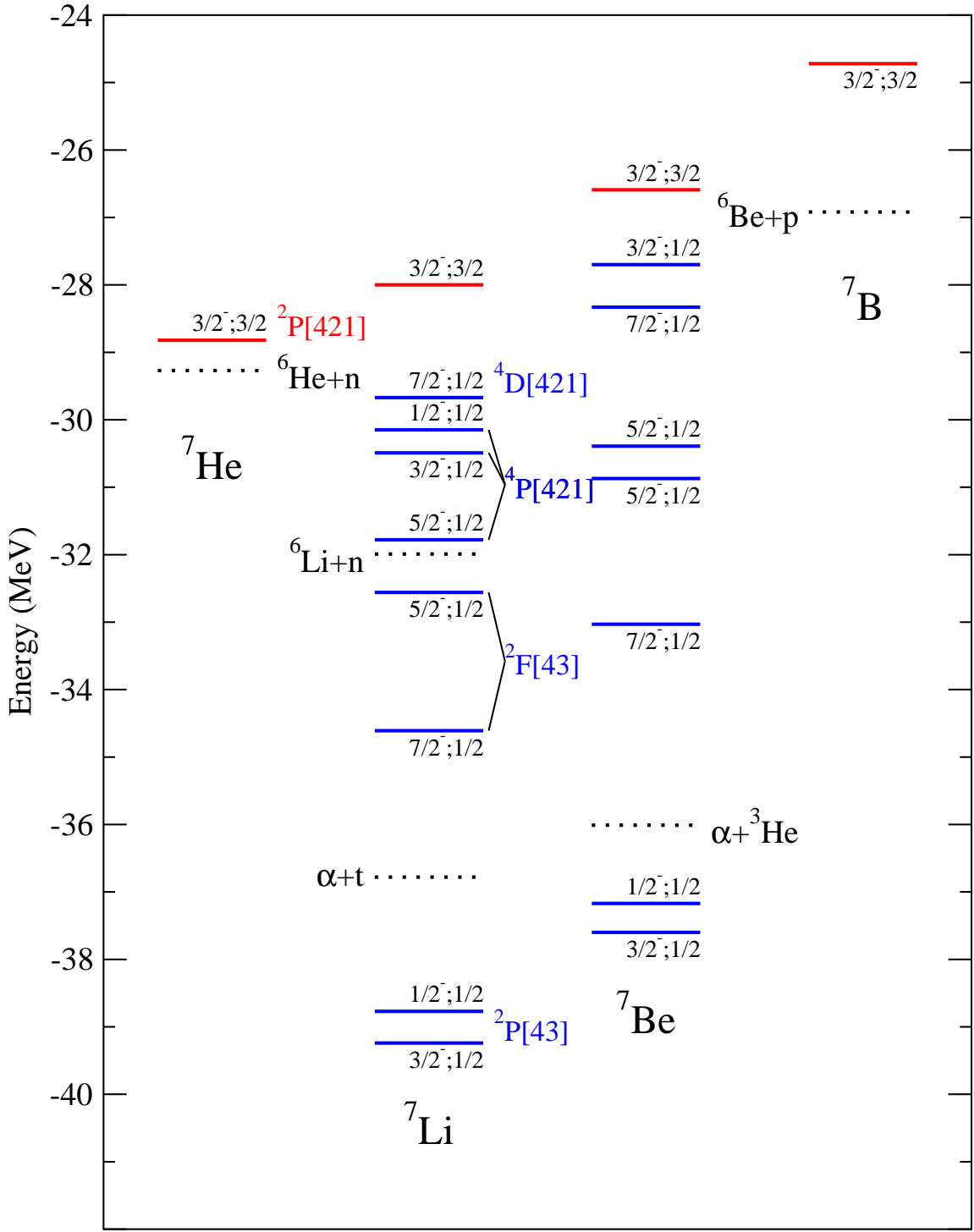}
\caption{(Color online) Experimental spectrum for $A=7$ nuclei; $T=1/2$ 
($T=3/2$) states are shown by blue (red) solid lines.}
\label{fig:a7}
\end{figure}

The $P_A(ST)$ and $E_{OPE}$ for $A=7$ nuclei are given in Table~\ref{tab:a7}.
Again we see that the s-shell core gives the same contribution as before,
and though there are now twelve sp pairs, they continue to give no net
contribution to $E_{OPE}$.  All the action is now in the three pp pairs.
In $^7$Li they can form part of a maximally symmetric $^2P$[43] or $^2F$[43]
state with energy contribution $-9C$ for a total $E_{OPE}=-27C$, which
equals the sum of alpha and triton energies; experimentally the ground 
state is 2.47 MeV below this sum, as seen in the experimental spectrum of 
Fig.~\ref{fig:a7}.  Again, much of the binding between clusters is apparently 
due to the $3N$ forces; GFMC calculations with AV18 alone produce a $^7$Li 
ground state only 0.3 MeV below the alpha-triton threshold \cite{PWC04}.

As labeled in the figure, the $^7$Li states are ordered according to our 
simple model, with $^2P$[43] and $^2F$[43] states lowest, followed by the 
$^4P$[421] states and the start of the $^4D$[421] states; VMC calculations 
confirm that these are by far the dominant components of the first eight 
states.  (The lowest five states in $^7$Be follow a similar pattern with
a Coulomb shift; the higher states may not be as well known experimentally.)
The $^4P$[421] states have a net $E_{OPE}=-21C$, the same as 
$^6$Li ground state, and they lie just above the threshold for breakup 
into $^6$Li+$n$.  The $^2P$[421] and $^4P$[421] $T=1/2$ states in $^7$Li 
have the same spatial symmetry but the former contain an admixture of the 
very repulsive $ST$=00 pairs, which pushes their energy up significantly,
and no corresponding experimental states have been identified.
In contrast, the $^2P$[421] $T=3/2$ ground state in $^7$He (and its isobaric
analogs) does not have any $ST$=00 pp pairs, and by our simple model has 
the same energy as the $^6$He ground state, which is about right.

\begin{figure}[ht!]
\centering
\includegraphics[width=6.0in]{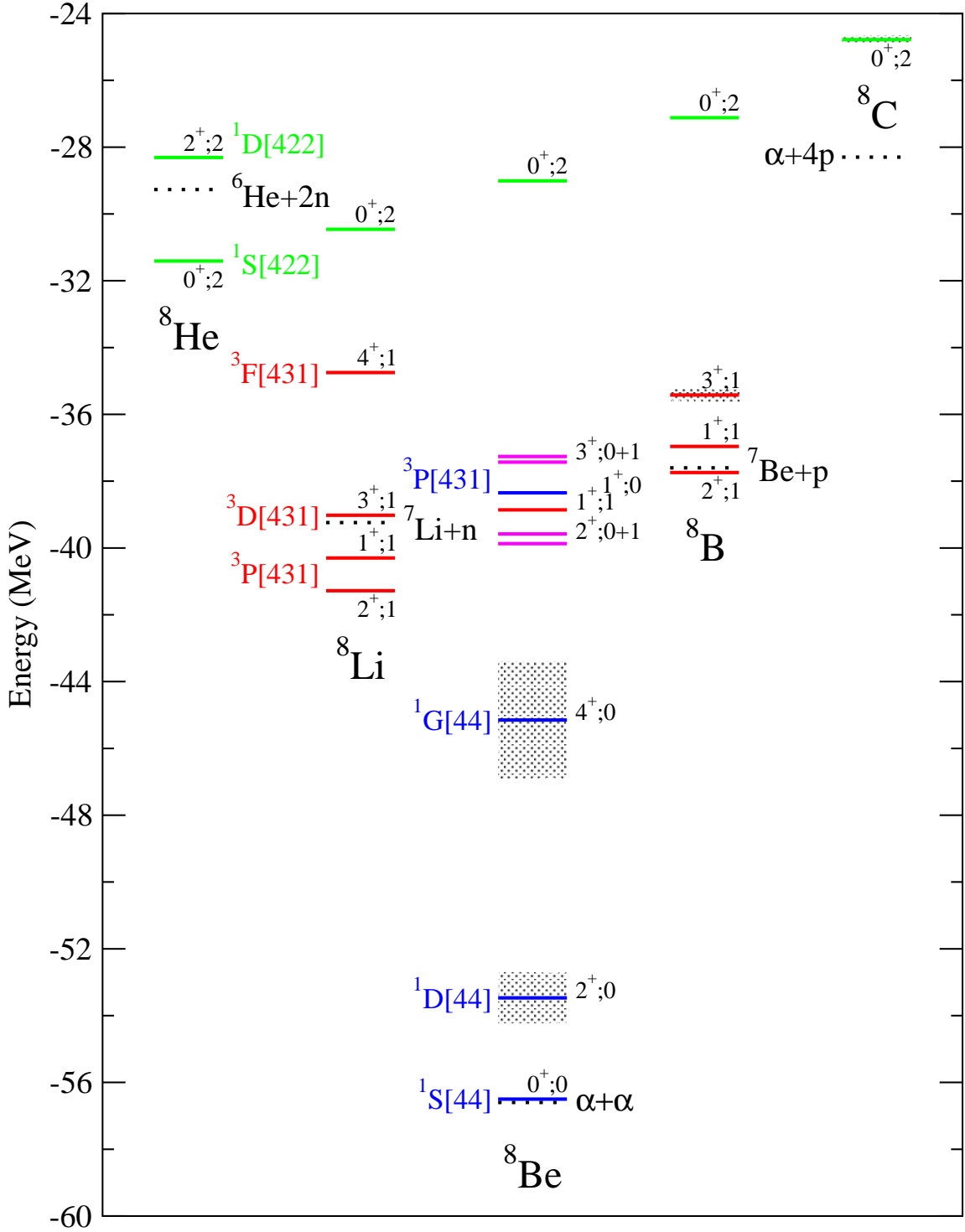}
\caption{(Color online) Experimental spectrum for $A=8$ nuclei: $T=0$, 1,
and 2 states are shown by blue, red, and green solid lines, respectively,
and isospin mixed states by magenta lines.}
\label{fig:a8}
\end{figure}

The $P_A(ST)$ and $E_{OPE}$ for $A=8$ nuclei are given in Tables~\ref{tab:be8}
and \ref{tab:li8he8}.  In the former we show the ss and sp pairs again
to remind us that each of the p-shell nuclei has an s-shell core 
contributing $-18C$ to $E_{OPE}$ and no contribution from the sp pairs.  
In $^8$Be the six pp pairs in the maximally symmetric [44] ground state 
effectively form a second alpha, so the total $E_{OPE}$ is 
$2\times(-18C)=-36C$, compared to $-28C$ for $^8$Li and $-24C$ for $^8$He 
ground states.  This is a fair representation of the spread in the experimental
spectrum, shown in Fig.~\ref{fig:a8} \cite{exp8910}.  The $^8$Be ground state 
is practically degenerate with the energy of two alphas, $^8$Li
is significantly less bound, but is a little more bound than $^7$Li ($-27C$),
and $^8$He is somewhat less bound, but below $^6$He and $^7$He (both $-21C$).
The increased binding for $^8$He is essentially due to the completion of
a second dineutron pair in its [422] symmetry ground state, which is worth
an additional $-3C$ in $E_{OPE}$.

The gap between the [44] and [431] symmetry states in $^8$Be has the large
value of $12C$, suggesting little mixing between them, and VMC calculations 
indicate the first $0^+$, $2^+$, and $4^+$ states are $\sim 99\%$ pure
symmetry [44].  By comparison, the small energy gap of $4C$ between the
[422] and [4211] symmetry states in $^8$He leads to mixed amplitudes of 
0.8~:~0.6 in its ground state \cite{PWC04}.  In $^8$Li there are both 
triplet and singlet states of symmetry [431], but the higher spin states 
fall lower in the spectrum because they avoid the repulsive $ST$=00 pairs; 
the same is true for the three different spin states of [4211] symmetry, 
and in $^8$Be for the two different spin states of [422] symmetry.

\begin{figure}[ht!]
\centering
\includegraphics[width=6.0in]{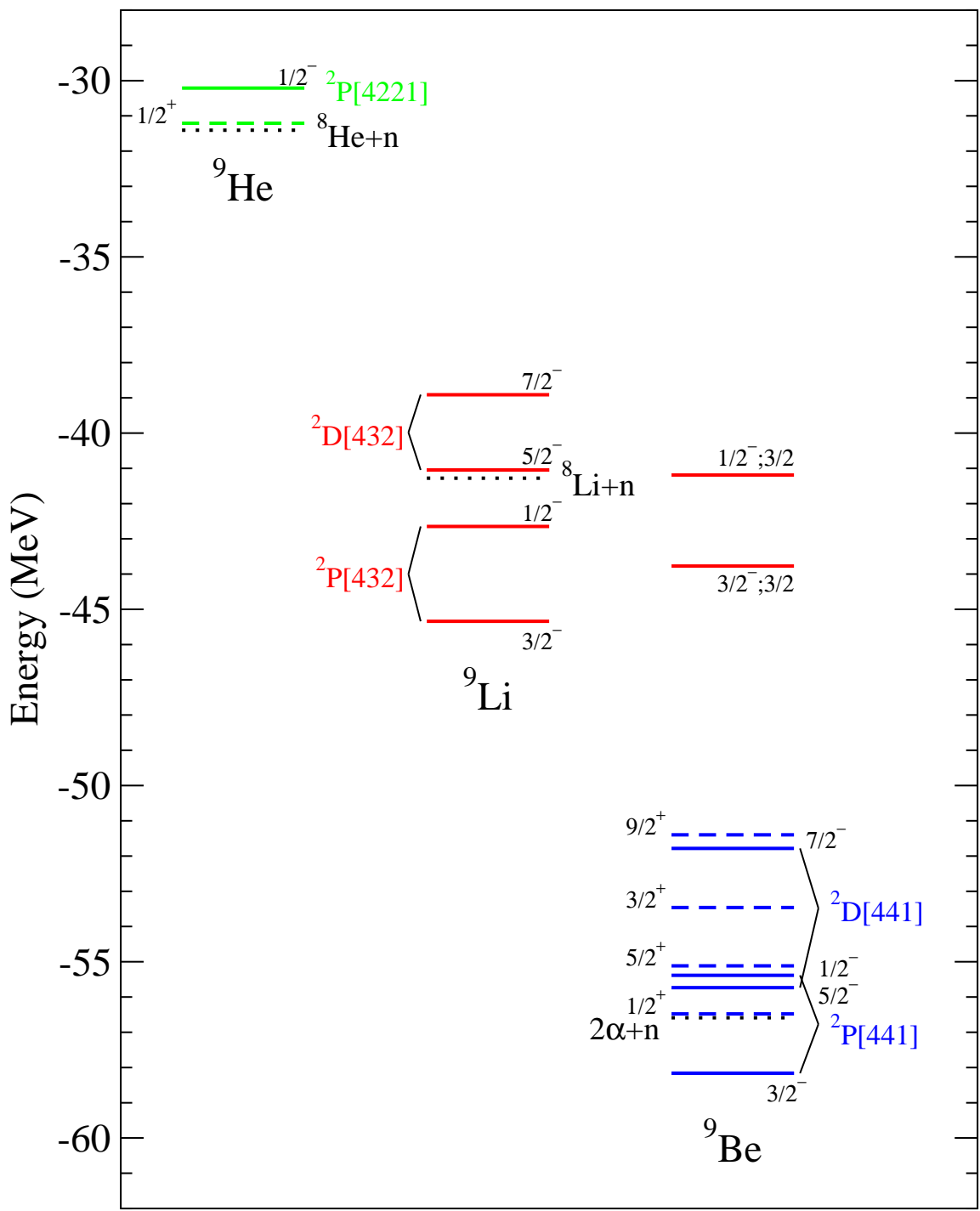}
\caption{(Color online) Experimental spectrum for $A=9$ nuclei: $T=1/2$,
3/2, and 5/2 states are shown by blue, red, and green lines; solid lines
denote natural-parity states and dashed lines unnatural-parity states.}
\label{fig:a9}
\end{figure}

The $P_A(ST)$ and $E_{OPE}$ in $A=9$ nuclei are given in 
Tables~\ref{tab:he9li9} and \ref{tab:be9}.  Adding in the contribution from
the s-shell core, the $E_{OPE}$ are $-36C$, $-30C$, and $-24C$ for the ground 
states of $^9$Be, $^9$Li, and $^9$He, respectively, which is again a very 
good approximation to the experimental spectrum shown in Fig.~\ref{fig:a9}
\cite{exp8910}.  The $^9$Be ground state is predicted to have the same energy
as $^8$Be or two alphas: the addition of one nucleon to the ground state of 
$^8$Be generates four new pp pairs in $^9$Be, but with just the right 
combination to add no additional binding to $E_{OPE}$.
Experimentally $^9$Be is bound with respect to the
threshold for alpha-alpha-neutron breakup by 1.57 MeV, which in turn is
0.10 MeV below the $^8$Be+$n$ threshold.  GFMC calculations indicate that 
the stability of the last neutron is again due to $3N$ forces: whereas the 
AV18/IL2 Hamiltonian gets $1.9 \pm 0.5$ MeV for the binding relative to 
$^8$Be, AV18 alone is stable by only $0.1 \pm 0.4$ MeV \cite{PVW02,PWC04}.  

The $^9$Li ground state is predicted to be somewhat more bound with respect 
to $^8$Li ($-30C$ compared to $-28C$) and experimentally it is stable by 
4.06 MeV.  On the other hand, $^9$He is predicted to be the same energy
as $^8$He (both $-24C$) because the last neutron is unpaired; experimentally 
the lowest natural-parity $1/2^-$ state is unbound by $\sim 1.2$ MeV.  
However, recent experiments indicate the lowest state in $^9$He is an unnatural 
positive-parity $1/2^+$ state just above threshold, and there are also 
many low-lying positive-parity states in $^9$Be, starting with a $1/2^+$
state just above the $2\alpha + n$ threshold.  These unnatural-parity 
states can be constructed by putting the last nucleon in an sd-shell orbital
outside the p-shell core.  As with the p-shell, the $E_{OPE}$ weight factor 
is such that there is no net interaction between an sd-shell nucleon and the 
core.  In the case of $^9$He and $^9$Be, our simple model suggests that the
the long-range part of the $N\!N$ potential does not care what orbital 
the last nucleon goes into; whether a p-shell or sd-shell orbital is more
stable depends on the residual shorter-range $N\!N$ interaction, the $3N$
interaction, and the kinetic energy cost.

There is a moderate gap of size $6C$ between the first and second symmetry 
states in $^9$Be, which is not as large as the gap in $^8$Be; consequently
the low-lying states are mostly [441] symmetry with relatively small 
admixtures of [432] components, but not as pure as the $^8$Be [44]
states \cite{PVW02}.  The smaller gap between symmetry states in $^9$Li
leads to more mixing of the [432] and [4311] components there.  VMC
diagonalizations also continue to show that for states of the same spatial
symmetry, those with higher spin lie lower in the spectrum, again due
to a smaller presence of repulsive $ST$=00 pairs.

\begin{figure}[ht!]
\centering
\includegraphics[width=6.0in]{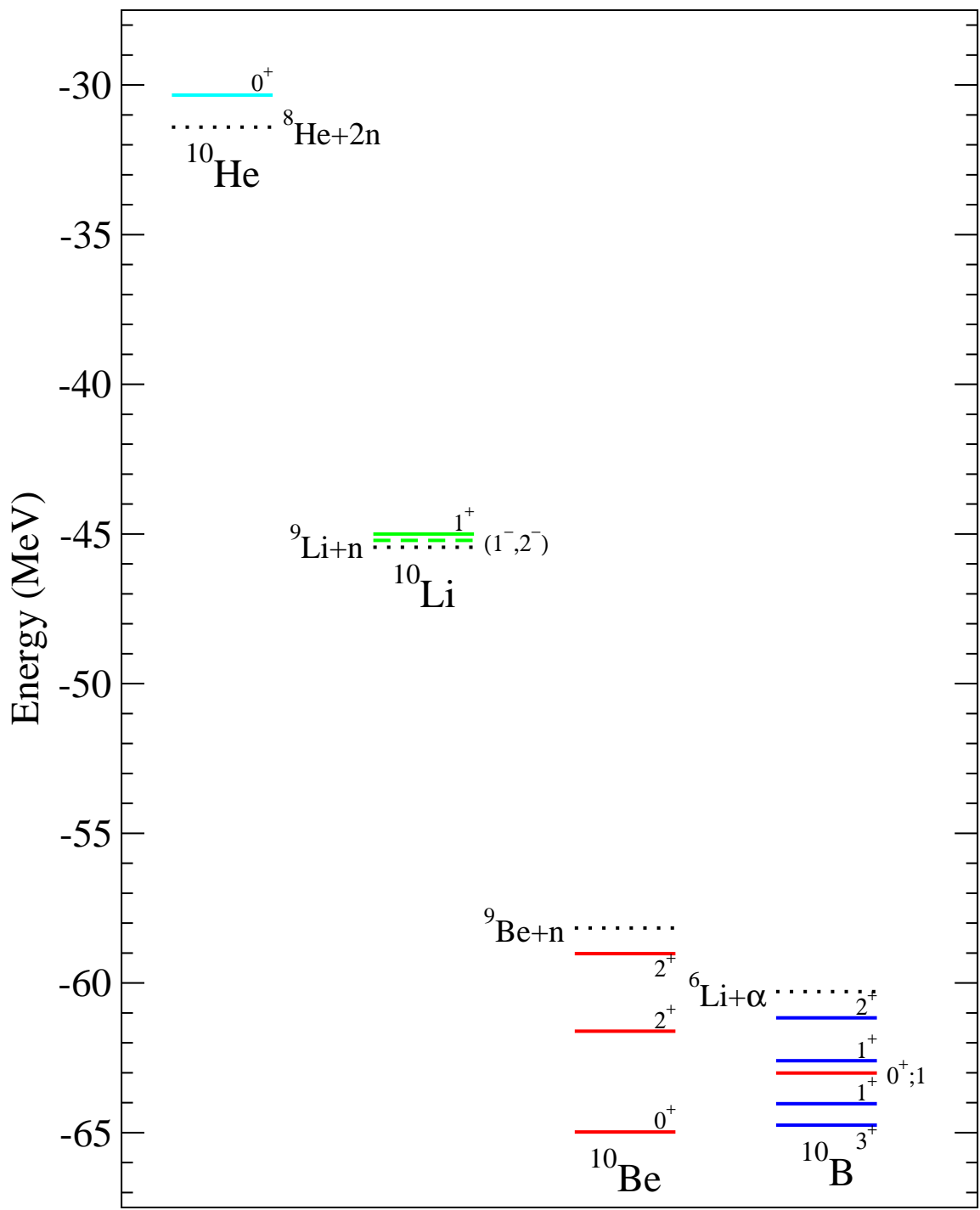}
\caption{(Color online) Simplified experimental spectrum for $A=10$ nuclei;
only stable natural parity states are shown for $^{10}$Be and $^{10}$B.}
\label{fig:a10}
\end{figure}

\begin{figure}[ht!]
\centering
\includegraphics[width=6.0in]{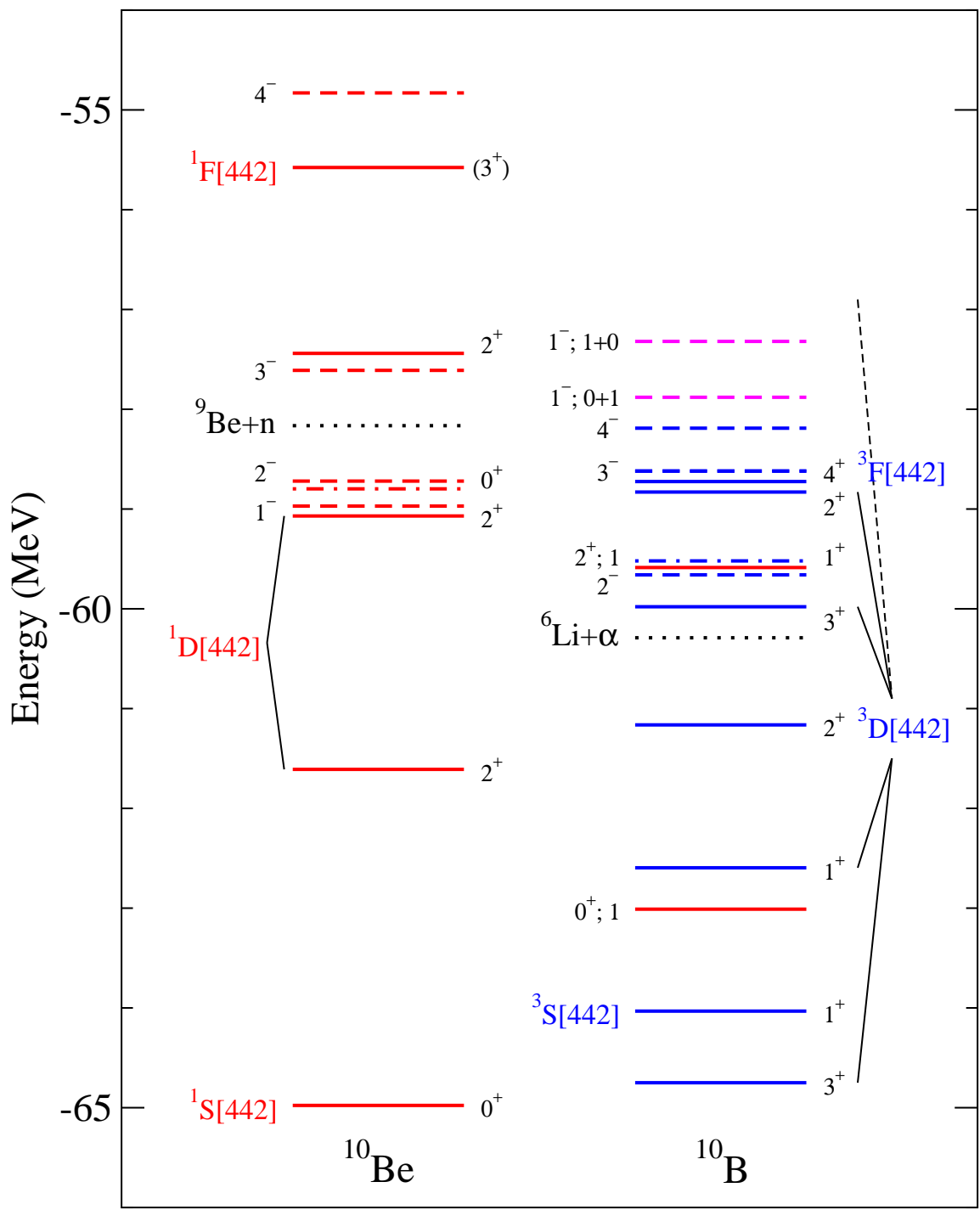}
\caption{(Color online) Detailed experimental spectrum for $^{10}$Be 
and $^{10}$B nuclei.}
\label{fig:b10}
\end{figure}

The $P_A(ST)$ and $E_{OPE}$ in $A=10$ nuclei are shown in 
Tables~\ref{tab:li10he10}, \ref{tab:be10}, and \ref{tab:b10}.  The $E_{OPE}$ 
are $-27C$, $-31C$, $-39C$, and $-39C$ for $^{10}$He, $^{10}$Li, $^{10}$Be,
and $^{10}$B, respectively.  Thus $^{10}$B and $^{10}$Be should have the 
same binding, which experimentally they do at 64.75 and 64.98 MeV as seen 
in Figs.~\ref{fig:a10} and \ref{fig:b10}.  They are predicted to be about 
4-5 MeV more bound than $^8$Be and $^9$Be, but experimentally it is more 
like 7-8 MeV.  On the other hand, the prediction for $^{10}$Li is that it 
should be a little more bound than $^9$Li, whereas it is unbound by about 
0.25 MeV.  Further, $^{10}$He should be bound by several MeV compared to 
$^8$He and $^9$He by the completion of another dineutron pair, whereas it 
is unbound by 1 MeV compared to $^8$He.  This could be an indication that 
$jj$-coupling is starting to be more appropriate as the neutron p-shell 
is completed \cite{CK67}, with this last pair of neutrons being a $p_{1/2}$ 
pair that joins at a noticeably higher energy than the first two dineutrons.
$^{10}$Li and $^{10}$He are the only two nuclei out of 27 in the p-shell 
(not counting isobaric analogs) that are falsely predicted to be stable by 
our simple model.

Comparing the different symmetry states in Table~\ref{tab:b10} for $^{10}$B, 
we see that the [442] components are $12C$ below the [4411], [433], etc.,
components, so there is very little admixture of the latter into the 
lowest-lying states.  However, in $^{10}$Be the gap between the ground state 
[442] symmetry and the [4411] and [433] components is only $4C$, so there 
is a moderate admixture into the ground state~\cite{PVW02}.  The clearest
signature for a state of these next spatial symmetries in the p-shell
would be a $1^+$ state in $^{10}$Be, expected at $\sim6$ MeV excitation,
but no such state has been observed.  However, unnatural-parity states
which involve an sd-shell intruder, are now low enough in the spectrum
for some of them to be particle-stable; discussion of these is deferred
to the next section.

In full GFMC calculations, the $3N$ force starts to make an especially large
impact by $A=10$ in that it starts to reorder some of the states from 
our simple expectations.  Naively we would expect the $^{10}$B spectrum
to be something like $^6$Li, with a $1^+$ $^3$S[442] ground state, and
a collection of $3^+$, $2^+$, and $1^+$ states above coming from the
spin-orbit splitting of the $^3$D[442] state.  The situation is complicated
by the fact that there are two linearly independent ways to construct an 
$L=2$ [442] symmetry state in the p-shell.  With AV18 only, the ground state
of $^{10}$B is in fact a $1^+$ state, but for AV18/IL2, the spin-orbit 
splitting of the $^3$D[442] states is large enough that one of the $3^+$ 
states is lowered to become the ground state \cite{PVW02}, as observed
experimentally.  
Similar results are obtained in NCSM calculations using the CD-Bonn or 
AV8$^\prime$ $N\!N$ potentials versus AV8$^\prime$ with the TM$^\prime(99)$ 
$3N$ potential added \cite{CNOV02,NO03}.  By comparison, $^{10}$Be behaves 
more like what we expect, with a $0^+$ ground state that is predominantly 
$^1$S[442] symmetry in character, while the next two $2^+$ excited states 
are dominated by the two $^1$D[442] symmetry combinations.

The complicated spectrum for $^{10}$B is shown in detail in Fig.~\ref{fig:b10}.
Based on GFMC calculations, the two $^3$D[442] triplets can be sorted by 
their quadrupole moments, $Q$.  One triplet with large positive $Q$ is widely
split and contains the ground state, the second $2^+$ near 6 MeV excitation,
and a predicted, but unobserved, fourth $1^+$ near 8 MeV.  The other
triplet has smaller negative $Q$ and is closely spaced, starting with the
second $1^+$ at 2 MeV excitation, followed by the first $2^+$ and second $3^+$.
The first $1^+$ is the $^3$S[442] state, while the third $1^+$ around 5 MeV 
excitation (marked by a dash-dot line in the figure) is believed to be a
$2\hbar\omega$ excitation.  Likewise, the second $0^+$ in $^{10}$Be near
6 MeV excitation is believed to be a $2\hbar\omega$ state.  These latter 
states will be discussed below with the unnatural-parity states.

The $A=10$ nuclei are the halfway points in the p-shell; moving further
up in the shell is comparable to removing particles from the filled
[4444] state of $^{16}$O.  The $A=11$ nuclei are the 5-hole complements 
of the 5- (p-shell) particle $A=9$ nuclei, $A=12$ nuclei are complements of 
$A=8$, etc.  For example, $^{11}$B is the complement of $^9$Be with the same 
allowed set of $^{2S+1}L$ components, except that [441] symmetry becomes 
[443], [432] becomes [4421], [4311] becomes [4331], and [4221] 
becomes [4322].  In like manner, $^{11}$Be is the complement of $^9$Li, and
$^{11}$Li is the complement of $^9$He.  Consequently we will not give 
tables for these heavier nuclei, except for $^{12}$C, which is of
particular interest as being at the present limits of GFMC and NCSM 
calculations with realistic forces \cite{P05,NVB00}, as well as being
an extremely popular experimental target.

Table~\ref{tab:c12} shows that $^{12}C$ has exactly the same $^{2S+1}L$ 
combinations as $^8$Be, with spatial symmetries augmented by an additional
[4] in the Young diagram.  The ground state will be a $^1S[444]$ $0^+$ state
with $E_{OPE}=3\times(-18)=-54C$ three times that of $^4$He.  
Experimentally $^{12}C$ is 7 MeV or 8\% more bound than three alphas.  
The distribution of the 28 pp pairs is such that sixteen average to give 
zero contribution to $E_{OPE}$ leaving twelve $ST=10$ and 01 pairs 
that are equivalent to two alphas in the p-shell.  In this simple 
model every time an alpha is formed in the p-shell, it effectively decouples 
from other nucleons in the p-shell.  There is again a large energy gap 
between the first and second symmetry components in $^{12}C$, so we
expect the ground and low-lying states to be predominantly [444] symmetry.
By contrast, we can predict that $^{12}$B and $^{12}$Be should have 
substantial mixing of different symmetries in their ground states.
This knowledge is of practical benefit for the quantum Monte Carlo 
calculations, where allowing for all the possible spatial symmetries 
in $A=12$ nuclei is computationally prohibitive at present. 

\begin{figure}[ht!]
\centering
\includegraphics[width=6.4in]{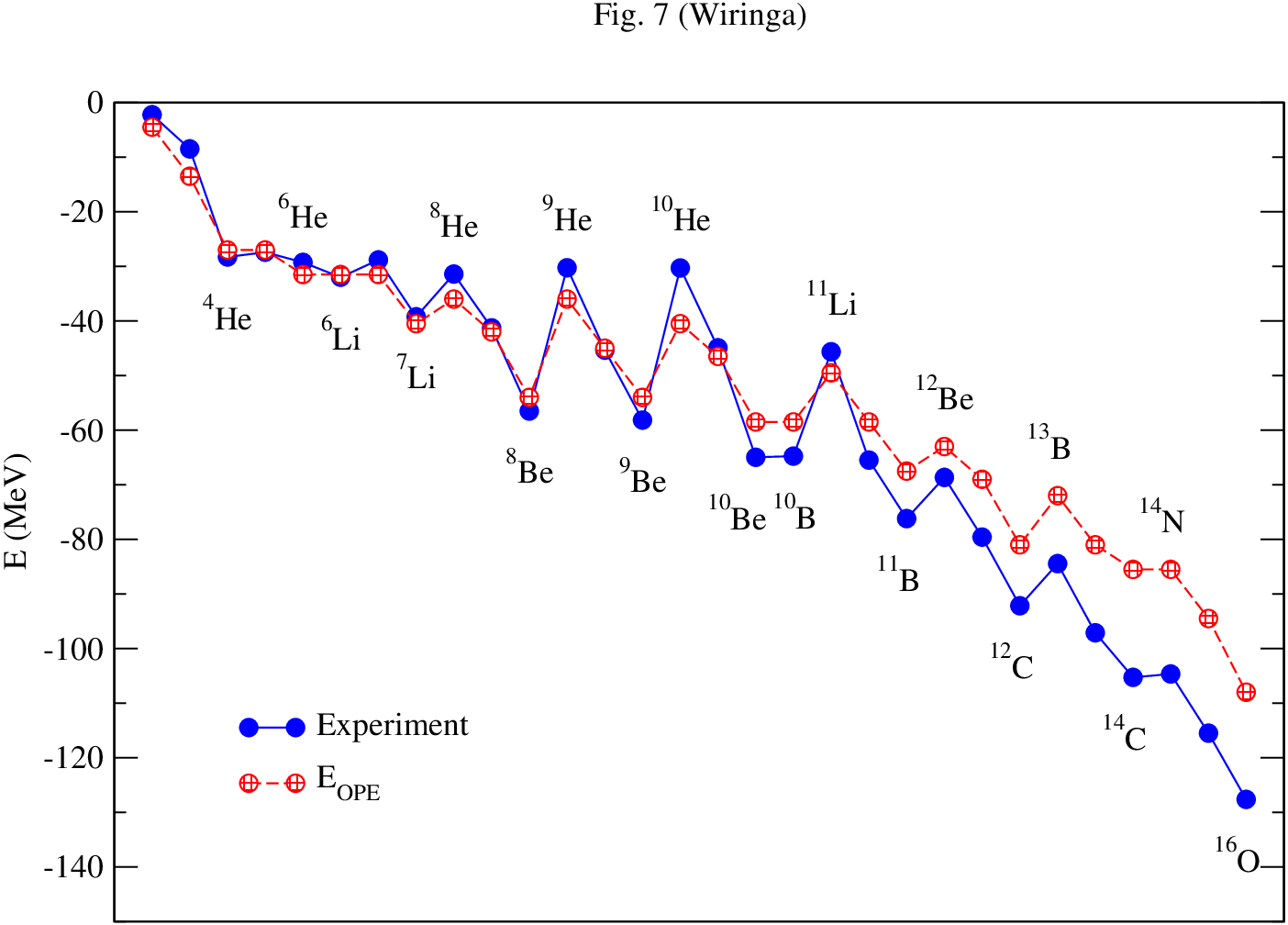}
\caption{(Color online) Ground state binding energies for $A \leq 16$ 
s- and p-shell nuclei.}
\label{fig:eope}
\end{figure}

The total energy for 30 s- and p-shell nuclei, ordered by increasing $A,Z$ 
but not including isobaric analogs, is plotted in Fig.~\ref{fig:eope}, where 
experiment and our $E_{OPE}$ are compared.  For this figure, we have set 
the coefficient in Eq.(\ref{eq:ope}) to be $C=1.5$ MeV.  The figure shows 
that up to $A\approx9$ the simple model works quite well, but then starts 
to underestimate the overall binding as $A$ increases.  Considering the 
necessity of including $3N$ forces in full GFMC and NCSM calculations to 
obtain the empirical binding energies, it is not surprising that a simple 
model based on pairwise forces will start to fail in this manner.  
Cohen and Kurath in their study of effective interactions for the p-shell 
\cite{CK65} found it difficult to fit all $6 \leq A \leq 16$ nuclei at the 
same time, and consequently made some models to fit only $A\geq8$ states.
They also found in their studies of spectroscopic factors \cite{CK67} that 
there is a gradual transition from $LS$-coupling to $jj$-coupling over the 
range $A$=9-14, and perhaps this transition is not unrelated to the 
increasing importance of $3N$ forces.

As mentioned above, our simple model also predicts $^{10}$He and 
$^{10}$Li to be definitely stable, when they are not.  In a number of other
cases, the model gives identical energies for neighboring nuclei, such
as $^{4,5}$He and $^{8,9}$Be, and cannot predict stability one way or the
other; this will be determined by finer details of the $N\!N$ and $3N$ 
forces and kinetic energy considerations.  Nevertheless, the simple formula 
reproduces the experimental trends fairly well.

The model naturally indicates that total energies are close to those
of summed $\alpha$, $t$, and $d$ subclusters, where applicable.  In fact,
the following energy relations hold for the maximally symmetric states
with $N \geq Z$:
\begin{eqnarray}
\label{eq:ma}
E(AZ=m\alpha) &=& E(AZ=m\alpha+n) = m E_\alpha \ , \\
E(AZ=m\alpha+2n) &=& E(AZ=m\alpha+3n) = m E_\alpha + E_{2n} \ , \\
E(AZ=m\alpha+4n) &=& E(AZ=m\alpha+5n) = m E_\alpha + 2 E_{2n} \ , \\
E(AZ=m\alpha+6n) &=& m E_\alpha + 3 E_{2n} \ , \\
E(AZ=m\alpha+d) &=& m E_\alpha + E_d \ , \\
E(AZ=m\alpha+t) &=& m E_\alpha + E_t \ , \\
\label{eq:li8}
E(AZ=m\alpha+t+n) &=& m E_\alpha + E_t - C \ , \\
E(AZ=m\alpha+t+2n) &=& m E_\alpha + E_t + E_{2n} \ , \\
\label{eq:li10}
E(AZ=m\alpha+t+3n) &=& m E_\alpha + E_t + E_{2n} - C \ , \\
E(AZ=m\alpha+t+4n) &=& m E_\alpha + E_t + 2 E_{2n} \ ,
\end{eqnarray}
where $m$ is the number of included alphas, and $E_{2n} = -3C$ is the energy of
a dineutron, which again in this simple model is equal to $E_d$.
Two non-trivial cases are Eqs.(\ref{eq:li8}) and (\ref{eq:li10}),
which in the p-shell would apply to $^{8,10}$Li and $^{12}$B.  The model
indicates there is a little extra binding, $C$ more than the sum of the
subclusters, on the addition of the last neutron.  This is an accurate
description of experiment for $^8$Li and $^{12}$B, but not for $^{10}$Li.

\section {Beyond the p-shell}

This simple model can be extended into the sd-shell, although the utility
of doing so will continue to diminish as $A$ increases.  Counting pairs
between the s-, p-, and sd-shells the contributions to $E_{OPE}$ again 
average out so there is no net interaction between the shells.  Then 
the progression from $^{16}$O to $^{17}$O, $^{18}$O, and $^{18}$F nuclei
is exactly analogous to the progression from $^4$He to $^5$He, $^6$He and
$^6$Li.  This is in rough accord with experiment, as is the prediction
that the multiple-alpha nuclei will continue as Eq.(\ref{eq:ma}).
However, it will also predict that $^{19}$F is definitely more bound
than $^{20}$O, which is not the case; among other things, Coulomb effects
are becoming important enough that they need to be treated explicitly.

However, the basic logic of this simple model may be applicable to 
sd-shell intruder states in the p-shell.  The intruder states in 
$A=10$ nuclei, where particle-stable intruders first occur, are
an example.  An interesting feature of the data is that the intruders 
in $^{10}$Be are ordered starting from the most bound level as
$1^-$, $2^-$, $3^-$, and $4^-$.  However, in $^{10}$B the order is
$2^-$, $3^-$, $4^-$, and then $1^-$.  The relative ordering of the $1^-$ 
and $2^-$ states in these nuclei can be understood in the following manner.
A major part of the $A=9$ ground state is $^8$Be($0^+$) plus an unpaired 
$1p$-shell nucleon ($1p_{3/2}$ orbital in $jj$ coupling) to which we add 
a spin-up or spin-down $2s$-shell nucleon.  Because these nucleons have 
on average no net interaction with the $^8$Be core, their pairwise 
interaction should dominate.  The $2^-$ state is a ``stretch'' state 
obtainable only if both spins of the pair are aligned, i.e., pure $S=1$, 
while the $1^-$ state will have some $S=0$ pair content.  In $^{10}$Be 
the last pair has $T=1$, so the $2^-$ state will be a $^3P$ pair, whereas 
the $1^-$ state will be partially a $^1S$ pair, which is more attractive 
--- hence the $1^-$ state should be lower in the spectrum.  In $^{10}$B 
the last pair has $T=0$, so the $2^-$ state will be a $^3S$ pair, 
whereas the $1^-$ state will have some admixture of $^1P$, which is 
(much) more repulsive --- hence the $1^-$ state will be (much) higher.
Preliminary VMC calculations with realistic interactions successfully
reproduce these level orderings and exhibit exactly this type of 
$S,T$ pair distribution.

One may also consider placing both last two nucleons outside the 
$^8$Be core into the sd-shell, either as a dineutron pair in $^{10}$Be
or a deuteron in $^{10}$B.  In our simple model, these would have the
same energy as the ground states, although in practice there would be
some reduction in binding due to the greater distance from the core of
these orbitals and the consequent overall loss of potential attraction.
In actual fact, the second $0^+$ in $^{10}$Be and third $1^+$ in $^{10}$B
(shown in Fig.~\ref{fig:b10} by dash-dot lines) are believed to be 
$2\hbar\omega$ excitations of this type.  These states pose an interesting 
challenge for both the GFMC and NCSM microscopic calculations.

The present simple model provides an interesting contrast to relativistic
mean-field theories, which commonly omit the pion with the argument that 
its contribution will spin-isospin average to zero in nuclear matter; 
such models have been applied to nuclei as light as $^{16}$O \cite{HS81}.
However, summing the expectation value of the OPE operator
${\bf\sigma}_{i}\cdot{\bf\sigma}_{j}~{\bf\tau}_{i}\cdot{\bf\tau}_{j}$ 
over all pairs we get a result that grows linearly with $A$ for the 
multiple-alpha nuclei.  In practice, the quantum Monte Carlo calculations 
with realistic forces find that OPE provides about 75\% of the net 
potential energy expectation value, although much of this comes from 
the tensor part of OPE \cite{PPWC01}.

\section {Conclusions}

We have presented a simple model for understanding the basic structure of
light nuclei.  It is based on counting the number of different $S,T$ pairs
that occur in a given nuclear state of specific spatial symmetry and 
multiplying by a numeric strength taken from one-pion exchange.  This
simple picture gives a good description for the growth of binding as
$A$ increases while showing saturation as the p-shell is reached.  It
explains the tendency of light nuclei to form $d$, $t$, and $\alpha$
subclusters, and a variety of features in the excitation spectra, 
including why for states of the same spatial symmetry, those of higher
$S$ are lower in the spectrum.  We hope this picture provides some
useful physical intuition.

\acknowledgments

I wish to thank D. F. Geesaman, D. Kurath, D. J. Millener, 
V. R. Pandharipande, and S. C. Pieper for valuable discussions.
This work is supported by the U. S. Department of Energy, 
Office of Nuclear Physics, under contract No. W-31-109-ENG-38.

\vfill
\newpage
\renewcommand{\baselinestretch}{1}
\begin{table}[ht!]
\caption{Pairs and OPE weights for $A$=2--5 nuclei.}
\begin{ruledtabular}
\begin{tabular}{lrrrrrr}
& \multicolumn{1} {l}{$^2$H}  & \multicolumn{1} {l}{$^3$H} 
& \multicolumn{1} {l}{$^4$He} & \multicolumn{2} {c}{$^5$He} \\ \cline{5-6}
& \multicolumn{1} {c}{$^3S$[2]} & \multicolumn{1} {c}{$^2S$[3]}
& \multicolumn{1} {c}{$^1S$[4]} & \multicolumn{2} {c}{$^2P$[41]} \\
$ST$ &    ss &    ss &    ss &    ss &    sp \\
\colrule
11   &       &       &       &       & \ninf \\
10   &     1 & \thrh &     3 &     3 & \thrf \\
01   &       & \thrh &     3 &     3 & \thrf \\
00   &       &       &       &       & \onef \\
\colrule
$P_A$&     1 &     3 &     6 &     6 &     4 \\
\colrule
$E_{OPE}$ &  $-$3C&  $-$9C& $-$18C& $-$18C&     0 \\
\end{tabular}
\end{ruledtabular}
\label{tab:a25}
\end{table}

\renewcommand{\baselinestretch}{1}
\begin{table}[ht!]
\caption{Ratio of kinetic to potential energy for $A \leq 12$ nuclei from
GFMC calculations with the AV18/IL2 Hamiltonian.}
\begin{ruledtabular}
\begin{tabular}{lrlr}
$^A$Z       & $R_{KV}$ &  $^A$Z       & $R_{KV}$ \\
\colrule
$^2$H       &   0.90   &  $^8$Li      &   0.79  \\
$^3$H       &   0.84   &  $^8$Be      &   0.76  \\
$^4$He      &   0.75   &  $^9$He      &   0.81  \\
$^6$He      &   0.78   &  $^9$Li      &   0.79  \\
$^6$Li      &   0.79   &  $^9$Be      &   0.77  \\
$^7$He      &   0.79   &  $^{10}$Be   &   0.77  \\
$^7$Li      &   0.78   &  $^{10}$B    &   0.77  \\
$^8$He      &   0.80   &  $^{12}$C    &   0.77  \\
\end{tabular}
\end{ruledtabular}
\label{tab:virial}
\end{table}

\begin{table}[ht!]
\caption{Pairs and OPE weights for $A$=6 nuclei. The $E_{OPE}$ weight does not
depend upon the total $L$ value, but we enumerate all possible values for a 
given spin and spatial symmetry combination; the allowed $T=0$ states are
given under the $^6$Li header, and the $T=1$ states under the $^6$He header.}
\begin{ruledtabular}
\begin{tabular}{lrrrrrr}
& & & \multicolumn{2} {c}{$^6$Li}      & \multicolumn{2} {c}{$^6$He} \\
\cline{4-5} \cline{6-7}
& & & \multicolumn{1} {c}{$^3SD$[42]}  & \multicolumn{1} {c}{$^1P$[411]}
    & \multicolumn{1} {c}{$^1SD$[42]}  & \multicolumn{1} {c}{$^3P$[411]} \\
$ST$ &    ss &    sp &    pp &   pp  &    pp &    pp \\
\colrule
11   &       & \ninh &       &       &       &     1 \\
10   &     3 & \thrh &     1 &       &       &       \\
01   &     3 & \thrh &       &       &     1 &       \\
00   &       & \oneh &       &     1 &       &       \\
\colrule
$P_A$&     6 &     8 &     1 &    1  &    1  &     1 \\
\colrule
$E_{OPE}$ & $-$18C&     0 &  $-$3C&  $+$9C&  $-$3C&  $+$1C 
\end{tabular}
\end{ruledtabular}
\label{tab:a6}
\end{table}

\begin{table}[ht!]
\caption{Pairs and OPE weights for $A$=7 nuclei.}
\begin{ruledtabular}
\begin{tabular}{lrrrrrrrr}
& & & \multicolumn{4} {c}{$^7$Li}       & \multicolumn{2} {c}{$^7$He} \\
\cline{4-7} \cline{8-9}
& & & \multicolumn{1} {c}{$^2PF$[43]}   & \multicolumn{1} {c}{$^4PD$[421]} 
    & \multicolumn{1} {c}{$^2PD$[421]}  & \multicolumn{1} {c}{$^2S$[4111]}
    & \multicolumn{1} {c}{$^2PD$[421]}  & \multicolumn{1} {c}{$^4S$[4111]} \\
$ST$ &    ss &    sp &    pp &    pp &    pp &    pp &   pp  &    pp \\
\colrule
11   &       & \twsf &       & \thrh & \thrf & \thrh & \thrh &     3 \\
10   &     3 & \ninf & \thrh & \thrh & \thrf &       &       &       \\
01   &     3 & \ninf & \thrh &       & \thrf &       & \thrh &       \\
00   &       & \thrf &       &       & \thrf & \thrh &       &       \\
\colrule
$P_A$&     6 &    12 &     3 &     3 &     3 &     3 &     3 &     3 \\
\colrule
$E_{OPE}$ & $-$18C&     0 &  $-$9C&  $-$3C&  $+$3C& $+$15C&  $-$3C&  $+$3C
\end{tabular}
\end{ruledtabular}
\label{tab:a7}
\end{table}

\begin{table}[ht!]
\caption{Pairs and OPE weights for $^8$Be.}
\begin{ruledtabular}
\begin{tabular}{lrrrrrrr}
& & & \multicolumn{5} {c}{$^8$Be} \\
\cline{4-8}
& & & \multicolumn{1} {c}{$^1SDG$[44]}  & \multicolumn{1} {c}{$^3PDF$[431]}
    & \multicolumn{1} {c}{$^5SD$[422]}  & \multicolumn{1} {c}{$^1SD$[422]} 
    & \multicolumn{1} {c}{$^3P$[4211]}   \\
$ST$ &    ss &    sp &    pp &    pp &    pp &    pp &    pp  \\
\colrule
11   &       &     9 &       & \thrh &     3 & \thrh & \fivh  \\
10   &     3 &     3 &     3 & \fivh &     3 & \thrh & \thrh  \\
01   &     3 &     3 &     3 & \thrh &       & \thrh & \oneh  \\
00   &       &     1 &       & \oneh &       & \thrh & \thrh  \\
\colrule
$P_A$&     6 &    16 &     6 &     6 &     6 &     6 &     6  \\
\colrule
$E_{OPE}$ & $-$18C& 0 &$-$18C&  $-$6C& $-$6C&  $+$6C& $+$10C
\end{tabular}
\end{ruledtabular}
\label{tab:be8}
\end{table}

\begin{table}[ht!]
\caption{Pairs and OPE weights for $^8$Li and $^8$He.}
\begin{ruledtabular}
\begin{tabular}{lrrrrrrrr}
& \multicolumn{6} {c}{$^8$Li}       & \multicolumn{2} {c}{$^8$He} \\
\cline{2-7} \cline{8-9}
& \multicolumn{1} {c}{$^3PDF$[431]} & \multicolumn{1} {c}{$^1PDF$[431]}
& \multicolumn{1} {c}{$^3SD$[422]}  & \multicolumn{1} {c}{$^5P$[4211]}
& \multicolumn{1} {c}{$^3P$[4211]}  & \multicolumn{1} {c}{$^1P$[4211]}
& \multicolumn{1} {c}{$^1SD$[422]}  & \multicolumn{1} {c}{$^3P$[4211]} \\
$ST$ &    pp &    pp &    pp &    pp &    pp &    pp &    pp &    pp \\
\colrule
11   &     2 & \thrh & \fivh &     4 &     3 & \fivh &     3 &     4 \\
10   &     2 & \thrh & \thrh &     2 &     1 & \oneh &       &       \\
01   &     2 & \fivh & \thrh &       &     1 & \thrh &     3 &     2 \\
00   &       & \oneh & \oneh &       &     1 & \thrh &       &       \\
\colrule
$P_A$&     6 &     6 &     6 &     6 &     6 &     6 &     6 &     6 \\
\colrule
$E_{OPE}$ &$-$10C&  $-$6C&  $-$2C&  $-$2C&  $+$6C& $+$10C&  $-$6C&  $-$2C
\end{tabular}
\end{ruledtabular}
\label{tab:li8he8}
\end{table}

\begin{table}[ht!]
\caption{Pairs and OPE weights for $^9$He and $^9$Li.}
\begin{ruledtabular}
\begin{tabular}{lrrrrrrrr}
& & & \multicolumn{1} {c}{$^9$He}   & \multicolumn{5} {c}{$^9$Li}     \\
\cline{5-9}
& & & \multicolumn{1} {c}{$^2P$[4221]} & \multicolumn{1} {c}{$^2PDF$[432]} 
& \multicolumn{1} {c}{$^4SD$[4311]}   & \multicolumn{1} {c}{$^2SD$[4311]} 
& \multicolumn{1} {c}{$^4P$[4221]}    & \multicolumn{1} {c}{$^2P$[4221]}  \\
$ST$ &    ss &    sp &    pp &    pp &    pp &    pp &    pp &    pp \\
\colrule
11   &       & \ftyf &     6 & \fitf &     5 & \svtf & \eleh & \nitf \\
10   &     3 & \fitf &       & \ninf & \fivh & \sevf &     2 & \fivf \\
01   &     3 & \fitf &     4 & \fitf & \fivh & \thtf &     2 & \elef \\
00   &       & \fivf &       & \onef &       & \thrf & \oneh & \fivf \\
\colrule
$P_A$&     6 &    20 &    10 &    10 &    10 &    10 &    10 &    10 \\
\colrule
$E_{OPE}$ & $-$18C& 0 & $-$6C& $-$12C& $-$10C&  $-$4C&  $-$1C&  $+$4C
\end{tabular}
\end{ruledtabular}
\label{tab:he9li9}
\end{table}

\begin{table}[ht!]
\caption{Pairs and OPE weights for $^9$Be.}
\begin{ruledtabular}
\begin{tabular}{lrrrrrrrr}
& \multicolumn{8} {c}{$^9$Be}        \\
\cline{2-9}
& \multicolumn{1} {c}{$^2PDFG$[441]} & \multicolumn{1} {c}{$^4PDF$[432]}
& \multicolumn{1} {c}{$^2PDF$[432]}  & \multicolumn{1} {c}{$^4SD$[4311]}
& \multicolumn{1} {c}{$^2SD$[4311]}  & \multicolumn{1} {c}{$^6P$[4221]}
& \multicolumn{1} {c}{$^4P$[4221]}   & \multicolumn{1} {c}{$^2P$[4221]} \\
$ST$ &    pp &    pp &    pp &    pp &    pp &    pp &    pp &    pp \\
\colrule
11   & \ninf & \fitf &     3 & \svtf & \sevh &     6 & \nitf &     4 \\
10   & \fitf & \fitf &     3 & \thtf & \fivh &     4 & \elef &     2 \\
01   & \fitf & \ninf &     3 & \sevf & \fivh &       & \fivf &     2 \\
00   & \onef & \onef &     1 & \thrf & \thrh &       & \fivf &     2 \\
\colrule
$P_A$&    10 &    10 &    10 &    10 &    10 &    10 &    10 &    10 \\
\colrule
$E_{OPE}$ &$-$18C& $-$12C&  $-$6C&  $-$4C&  $+$2C&  $-$6C&  $+$4C& $+$10C
\end{tabular}
\end{ruledtabular}
\label{tab:be9}
\end{table}

\begin{table}[ht!]
\caption{Pairs and OPE weights for $^{10}$He and $^{10}$Li states.}
\begin{ruledtabular}
\begin{tabular}{lrrrrrr}
& & & \multicolumn{1} {c}{$^{10}$He}   & \multicolumn{3} {c}{$^{10}$Li}  \\
\cline{5-7}
& & & \multicolumn{1} {c}{$^1S$[4222]} & \multicolumn{1} {c}{$^3PD$[4321]}  
&     \multicolumn{1} {c}{$^1PD$[4321]} & \multicolumn{1} {c}{$^3S$[4222]}  \\
$ST$ &    ss &    sp &    pp &    pp &    pp &    pp \\
\colrule
11   &       & \twsh & \twnf & \twsf &     8 &     9 \\
10   &     3 & \ninh & \elef & \ninf &     2 &       \\
01   &     3 & \ninh & \nitf & \twof &     4 &     6 \\
00   &       & \thrh & \onef & \thrf &     1 &       \\
\colrule
$P_A$&     6 &    24 &    15 &    15 &    15 &    15 \\
\colrule
$E_{OPE}$ & $-$18C& 0 & $-$9C& $-$13C&  $-$9C&  $-$1C
\end{tabular}
\end{ruledtabular}
\label{tab:li10he10}
\end{table}

\begin{table}[ht!]
\caption{Pairs and OPE weights for $^{10}$Be states.}
\begin{ruledtabular}
\begin{tabular}{lrrrrrrrr}
& \multicolumn{8} {c}{$^{10}$Be}     \\
\cline{2-9}
& \multicolumn{1} {c}{$^1SD^*FG$[442]} & \multicolumn{1} {c}{$^3PF$[4411]}
& \multicolumn{1} {c}{$^3PF$[433]}   & \multicolumn{1} {c}{$^5PD$[4321]}
& \multicolumn{1} {c}{$^3P^*D^*$[4321]}  & \multicolumn{1} {c}{$^1PD$[4321]}
& \multicolumn{1} {c}{$^5S$[4222]}   & \multicolumn{1} {c}{$^1S$[4222]} \\
$ST$ &    pp &    pp &    pp &    pp &    pp &    pp &    pp &    pp \\
\colrule
11   & \ninh & \eleh & \eleh & \twnf & \twff & \twtf &     8 & \thth \\
10   & \ninh & \ninh & \ninh & \nitf & \fitf & \thtf &     4 & \fivh \\
01   & \eleh & \ninh & \ninh & \elef & \fitf & \svtf &     2 & \sevh \\
00   & \oneh & \oneh & \oneh & \onef & \fivf & \sevf &     1 & \fivh \\
\colrule
$P_A$&    15 &    15 &    15 &    15 &    15 &    15 &    15 &    15 \\
\colrule
$E_{OPE}$ &$-$21C& $-$17C& $-$17C& $-$13C&  $-$5C&  $-$1C&  $-$1C& $+$11C
\end{tabular}
\end{ruledtabular}
$^*$ denotes two linearly-independent combinations
\label{tab:be10}
\end{table}

\begin{table}[ht!]
\caption{Pairs and OPE weights for $^{10}$B states.}
\begin{ruledtabular}
\begin{tabular}{lrrrrrrr}
& \multicolumn{7} {c}{$^{10}$B}      \\
\cline{2-8}
& \multicolumn{1} {c}{$^3SD^*FG$[442]} & \multicolumn{1} {c}{$^1PF$[4411]}
& \multicolumn{1} {c}{$^1PF$[433]}   & \multicolumn{1} {c}{$^5PD$[4321]}
& \multicolumn{1} {c}{$^3PD$[4321]}  & \multicolumn{1} {c}{$^7S$[4222]}
& \multicolumn{1} {c}{$^3S$[4222]}   \\
$ST$ &    pp &    pp &    pp &    pp &    pp &    pp &    pp \\
\colrule
11   & \ninh & \ninh & \ninh & \twsf & \twtf &     9 & \thth \\
10   & \eleh & \ninh & \ninh & \twof & \svtf &     6 & \sevh \\
01   & \ninh & \ninh & \ninh & \ninf & \thtf &       & \fivh \\
00   & \oneh & \thrh & \thrh & \thrf & \sevf &       & \fivh \\
\colrule
$P_A$&    15 &    15 &    15 &    15 &    15 &    15 &    15 \\
\colrule
$E_{OPE}$ &$-$21C&  $-$9C&  $-$9C&  $-$9C&  $-$1C&  $-$9C& $+$11C
\end{tabular}
\end{ruledtabular}
$^*$ denotes two linearly-independent combinations
\label{tab:b10}
\end{table}

\begin{table}[ht!]
\caption{Pairs and OPE weights for $^{12}$C states.}
\begin{ruledtabular}
\begin{tabular}{lrrrrrrr}
& & & \multicolumn{5} {c}{$^{12}$C} \\
\cline{4-8}
& & & \multicolumn{1} {c}{$^1SDG$[444]}  & \multicolumn{1} {c}{$^3PDF$[4431]}
    & \multicolumn{1} {c}{$^5SD$[4422]}  & \multicolumn{1} {c}{$^1SD$[4422]}
    & \multicolumn{1} {c}{$^3P$[4332]}   \\
$ST$      &    ss &    sp &    pp &    pp &    pp &    pp &    pp  \\
\colrule
11        &       &    18 &     9 & \twoh &    12 & \twoh & \twth  \\
10        &     3 &     6 &     9 & \svth &     9 & \fith & \fith  \\
01        &     3 &     6 &     9 & \fith &     6 & \fith & \thth  \\
00        &       &     2 &     1 & \thrh &     1 & \fivh & \fivh  \\
\colrule
$P_A$     &     6 &    32 &    28 &    28 &    28 &    28 &    28  \\
\colrule
$E_{OPE}$ & $-$18C&     0 & $-$36C& $-$24C& $-$24C& $-$12C&  $-$8C
\end{tabular}
\end{ruledtabular}
\label{tab:c12}
\end{table}

\newpage

\newpage

\newpage

\newpage

\newpage

\newpage

\newpage

\end{document}